\documentclass[a4paper, 10pt, conference]{ieeeconf}      

\IEEEoverridecommandlockouts                              
\usepackage{amsmath} 
\usepackage{amssymb}  

\newcommand{\e}{\varepsilon}
\def\a{\alpha}

\def\c{\chi}
\def\d{\delta}
\def\et{\widetilde{\varepsilon}}
\def\p{\psi}
\def\m{\mu}
\newtheorem{theorem}{\bf Theorem}
\newtheorem{lemma}{\bf Lemma}

\title{\LARGE \bf
Limit points of the monotonic schemes
}

\author{Julien Salomon
\thanks{J. Salomon is with Laboratoire Jacques-Louis Lions, Universit{\'e} Pierre et Marie Curie,
  Bo{\^i}te courrier 187, 75252 Paris Cedex 05, France
  {\tt\small Salomon@ann.jussieu.fr}}}

\begin{document}

\maketitle
\thispagestyle{empty}
\pagestyle{empty}

\begin{abstract}
Many numerical simulations in quantum (bilinear) control use the
monotonically convergent algorithms of Krotov (introduced by Tannor in
\cite{tannor}), Zhu \& Rabitz (\cite{zhu}) or the general form of Maday \& Turinici (\cite{jcp}). This paper
presents an analysis of the limit set of controls provided by
these algorithms and a proof of convergence in a particular case.
\end{abstract}

\section{INTRODUCTION}
The control of quantum phenomena is a topic that has been (and is
still) a source of many interesting challenges not only to physics and chemistry
but also to the mathematics and applied mathematics communities (\cite{br,book}).
At the level of the experiments, laser control of complex molecular systems is becoming feasible,
especially since the introduction (\cite{judson,shi1}) of closed loop
laboratory learning techniques and their successful implementation
(\cite{scienceref1,hbref1,hbref2,hbref4,hbref5,hbref6}). \\
On the other hand, at the level of the numerical simulations, the introduction of the
monotonically convergent algorithms by Zhu \& Rabitz (\cite{zhu}) that
extend an algorithm due to Krotov (\cite{tannor}) has allowed a
considerable progress and made possible further investigations in this
area. Recently, a general class of monotonically convergent algorithms has
been proposed (\cite{jcp}) and a relevant time discretization has been
developed (\cite{seville}).\\
However, no general analysis to explain in depth the convergence of
these algorithms is available to date. In an attempt to fill this gap,
this paper presents some results on the set of the controls provided
by monotonic algorithms.\\Note that, among others, this question was raised in \cite{grivopoulos}, 
but a wrong statement about the Cauchy character of the sequence is made
that makes the proof not working as stated; the proof is more involved as
explained in what follows.\\
The balance of the paper is as follows: the necessary background and
definitions of the quantum control settings are given in Section II,
properties of the monotonic sequences are presented in Section III,
followed by the properties of the limit set in Section IV. Further
results in a particular case are given in Section V
and concluding remarks are presented in Section VI.

\section{QUANTUM OPTIMAL CONTROL AND MONOTONIC SCHEMES}
\subsection{Cost functional and Euler-Lagrange Equations}
Consider a quantum system prepared in an initial state $\psi_0$ and whose
dynamics is characterized by its internal Hamiltonian $H$. By assumption
this Hamiltonian does not give rise to an appropriate evolution and an
external interaction is introduced in order to obtain the desired final property. This
interaction is taken here as an electric field with
time-dependent amplitude $\varepsilon(t)$  that influences the system through a
time-independent dipole moment operator $\mu$. The new Hamiltonian
$H-\mu \varepsilon(t)$ gives rise to the equations (we work in atomic
units i.e. $\hbar=1$):

\begin{eqnarray} \label{eq2}
& \ & i \frac{\partial}{\partial t } \psi(x,t) =
H \psi(x,t) -  \mu(x) \varepsilon(t)\psi(x,t)\nonumber  \\
 & \ &  \psi (x,t=0) = \psi_0 (x),  \nonumber
\end{eqnarray}

\noindent where we denote by $x$ the relevant spatial coordinates. These
equations hold on $\Omega=\mathbb{R}^N$ but for numerical tests we will consider that $x$
belongs to an interval $\Omega=[0,L]$ and that $\psi(0,t)=\psi(L,t)=0$, for a large
enough real number $L$ and any $t$ in $\mathbb{R}$. This approach is justified by physical reasons
since wave functions are generally localized in a space interval.

The optimal control framework is then used to find a suitable evolution of $\varepsilon(t)$. The goal that the final state
$\psi(T)$ has prescribed properties is expressed by the introduction of
a cost functional $J$ to be maximized. This cost functional also includes
a contribution that penalizes undesirable
effects. One simple example of such a cost functional is:

\begin{equation} \label{p6-1}
J(\varepsilon) = \langle \psi(T) | O | \psi(T)
\rangle -  \alpha \int_0^T \varepsilon^2 (t) dt,
\end{equation}
where $\alpha  > 0$ is a parameter (it may also depend on time
cf. \cite{hornung}, \cite{innsbruck})  and
$O$ is an observable operator that encodes the goal:
the larger the value $ \langle \psi(T) | O | \psi(T)  \rangle  $ is, the
better the
control objectives are met (here and in what follows
we use the convention that for any functions $f$ and $g$ and
any operator $F$: $  \langle f | F | g  \rangle  =
 \int \overline{f(x)}Fg(x)dx $.
Note that, in general, achieving the maximal possible value of
$\langle \psi(T) | O | \psi(T)  \rangle$ is at the price of a large laser
influence
$\int_0^T \varepsilon^2 (t) dt $ ;
the optimum
evolution will therefore strike a balance between using a low
laser fluence while simultaneously maximizing
the desired observable.

At the maximum of the cost functional $J(\varepsilon)$, the Euler-Lagrange critical point equations are satisfied ;
a standard way to write these equations is to use
 a Lagrange multiplier  $\chi(x,t)$ called
{\it adjoint state}. The following
critical point equations
are thus obtained (\cite{zhu}):
\begin{eqnarray} \label{el1}
& \! & \left\{
\begin{array}{l}
{ i \frac{\partial}{\partial t}} \psi(x,t) =
( H- \varepsilon(t)\mu) \psi(x,t) \\
\psi(x,t=0) = \psi_0(x)
\end{array} \right.
\\ \label{el2}
& \! & \left\{
\begin{array}{l}
{ i \frac{\partial}{\partial t}} \chi(x,t) =
(H -\varepsilon(t)\mu) \chi(x,t) \label{el3} \\
\chi (x,t=T) = O \psi(x,T)
\end{array} \right.
\\ \label{el4}
& \! & \alpha \varepsilon(t) = - \textrm{Im}  \langle \chi(t) | \mu | \psi(t) \rangle   .
\end{eqnarray}
From now on, $\p(t)$ and $\c(t)$ will represent two functions of the
Hilbert space $\mathbb{L}^2(\Omega; \mathbb{C})$ for almost all $t$ in
$[0,T]$.

\subsection{Definition of the monotonic schemes}

Efficient strategies for solving in practice the
critical point equations (\ref{el1})-(\ref{el4}) are represented by  the
monotonically convergent algorithms (\cite{zhu,tannor,jcp}) that are
guaranteed to improve the cost functional $J$ at each iteration.
In the formulation proposed in \cite{jcp}, the monotonic algorithms are
described by the resolution of the following equations at step $k$:

\begin{eqnarray}
& \! &
\left\{\begin{array}{l}\label{monotonic}
{ i\frac{\partial
}{\partial t}}\p^{k}(t)=(H-\e^k(t)\m)\p^k(t)
\\ \p^k(x,t=0)=\p_0(x)
\end{array} \right.\\ \label{monotonic2}
& \! & \e^k(t)=(1-\d)\et^{k-1}(t)-\frac{\d}{\a}\textrm{Im}\langle \c^{k-1}(t)|\m|\p^k(t)\rangle\\
& \! & \left\{\begin{array}{l}\label{monotonic3}
{i\frac{\partial}{\partial t}}
    \c^{k}(t)=(H-\et^k(t)\m)\c^k(t)
\\ \c^k(x,t=T)=O\p^k(x,T)
\end{array} \right.\\
& \! & \et^k(t)=(1-\eta)\e^{k}(t)-\frac{\eta}{\a}\textrm{Im}\langle
\c^{k}(t)|\m|\p^k(t)\rangle .\label{monotonic4}
\end{eqnarray}
where $\delta$ and $\eta$ are two real parameters.

The most important property of this algorithm is given in the following
theorem (\cite{jcp}):\\
\begin{theorem}
Suppose $O$ is a self-adjoint positive semi-definite operator. Then,
for any $\eta, \delta \in [0,2]$ the algorithm given in
Eqns.~(\ref{monotonic})-(\ref{monotonic2}) converges monotonically in the sense
that:
\begin{equation}\label{monoprop} J(\e^{k+1}) \geq J(\e^k).\end{equation}
\end{theorem}



\section{PROPERTIES OF THE SEQUENCE $(\e^k)_{k}$, $(\et^k)_{k}$}
We first prove that $(\e^k)_k$ and $(\et^k)_k$
defined in (\ref{monotonic}) and in (\ref{monotonic2}) are bounded. We
then prove that every weakly convergent subsequence is strongly convergent.
In the following, $||.||$ represents the norm of
$\mathbb{L}^2(\Omega,\mathbb{C})$, whereas $\|.\|_2$ represents the
norm
of $\mathbb{L}^2([0,T]; \mathbb{R})$. The scalar product in $\mathbb{L}^2([0,T];\mathbb{R})$ will be denoted by $<.,.>$.
\subsection{Bound for the sequences}

We suppose from now on that $O$ and $\m$ are bounded operators and we
denote by  $||O||_*$, $||\mu||_*$ their norms.
\begin{theorem}\label{borne}
There exists $M>0$ such that, for all $k>0$, the solutions $\e^k$,
$\et^k$ of (\ref{monotonic}-\ref{monotonic4}) verify:
\begin{eqnarray}
  \forall t\in[0,T],\ |\varepsilon^{k+1}(t)| \leq M,\
  |\et^{k+1}(t)|\leq M .\nonumber
\end{eqnarray}
\end{theorem}
\vspace{.5cm}
\begin{proof}
 Define $M$ by:
\begin{equation}\label{defM}
 M\!\!=\!\! \max (\|\e^0\|_2,\|\et^0\|_2,\max (1,\frac{\d}{2-\d},\frac{\eta}{2-\eta})\frac{||O||_*||\m||_*} {\a})
,\end{equation} and assume that it has
been proven that $\| \et^{k-1} \|_2 \leq  M, \ \| \e^{k-1}\|_2 \leq  M$.
Since we also have:
\begin{equation}\nonumber
\|\e^k\|_2 \leq |1-\d| M+\frac{\d}{\a}\| \ t\mapsto|\textrm{Im} \langle \c^{k-1}(t)|\m|\p^k(t)\rangle| \
\|_2, \end{equation}
the Cauchy-Schwartz inequality yields:
\begin{eqnarray}\nonumber
  |\langle \c^{k-1}(t)|\m|\p^k(t)\rangle | &\leq& ||\c^{k-1}(t)||.||\m(\p^k(t))||\\
& \leq& ||\m||_*.||\c^{k-1}(t)||.||\p^k(t)||.\nonumber
\end{eqnarray}
We then use the following equalities and bounds on state and adjoint
state:
\begin{eqnarray}\nonumber
\forall t, ||\p^k(t)||&=&1,\\
\forall t, ||\c^{k-1}(t)||&=&||\c^{k-1}(T)||=||O(\p^{k-1}(T))||\nonumber\\ &\leq&
  ||O||_*.||\p^{k-1}(T)||=||O||_*\nonumber
\end{eqnarray}
to obtain the estimate:
\begin{equation}\nonumber
 \|\e^k\|_2 \leq  |1-\d| M+\d
   \frac{||O||_*.||\m||_*}{\a} .
\end{equation}
If $\d \leq 1$, then the definition (\ref{defM}) yields $  \frac{||O||_*.||\m||_*} {\a}<M
$ and then: $\|\e^k\|_2 \leq  |1-\d|M+\d M = M\ ,$
and if $\d >  1$ then $
\frac{\d}{2-\d}\frac{||O||_*||\m||_*}{\a}<M$ and in this
case:
$\|\e^k\|_2 \leq |1-\d|M+\d \frac{2-\d}{\d}M=(\d-1)M+(2-\d)M= M.$\\
A similar proof leads to the same estimate for $\et^k$.
\end{proof}
\subsection{Weak convergence of subsequences}
\subsubsection{Extraction of a weakly convergent subsequence}
Because $\e^k$ is bounded in the Hilbert space
$\mathbb{L}^2([0,T];\mathbb{R})$, there exists a weakly convergent subsequence that will be
denoted by $(\e^{k_n})_n$. Let $\e$ be the weak limit associated to $(\e^{k_n})_n$.

\subsubsection{Limits of $(\e^{k_n+1}-\e^{k_n})_n $ and $(\et^{k_n+1}-\et^{k_n})_n $}

The sequence $J(\e^k)$ is bounded since $|J(\e^k)|\leq
\|O\|+M$. It has also been proven (\cite{jcp}) that:

\begin{eqnarray}J(\e^{k+1})\!\!-\!\!J(\e^k)\!\!\!\! &\!\!=\!\!&\!\!\!\!\langle\p^{k+1}(T)-\p^k(T)|O|\p^{k+1}(T)-\p^k(T)\rangle\nonumber \\
\!\!&\!\!\!\!&\!\!
+\int_0^T(\frac{2}{\d}-1)(\e^{k+1}(t)-\et^k(t))^2dt\nonumber \\
\!\!&\!\!\!\!&\!\!+\int_0^T(\frac{2}{\eta}-1)(\et^k(t)-\e^k(t))^2dt,\nonumber\end{eqnarray}
which gives after summation:
\begin{eqnarray}J(\e^{N})\!\!-\!\!J(\e^0)\!\!\!\! &\!\!\!\! =\!\!\!\!
  &\!\!\!\!\!\! \sum_0^{N-1}\!\!\langle\p^{k+1}(T)-\p^k(T)|O|\p^{k+1}(T)-\p^k(T)\rangle
  \nonumber \\ &&
  +\int_0^T(\frac{2}{\d}-1)\sum_0^{N-1}(\e^{k+1}(t)-\et^k(t))^2dt\nonumber
 \\&&+\int_0^T(\frac{2}{\eta}-1)\sum_0^{N-1}(\et^k(t)-\e^k(t))^2dt.\nonumber\end{eqnarray}
Thus the series $\sum_0^{N-1}\|\e^{k+1}-\et^k\|_2^2$ and
  $\sum_0^{N-1}\|\et^k-\e^k\|_2^2 $ converge and we
  deduce that:
\begin{equation}\label{limite}
\lim_n\|\e^{k_n+1}-\e^{k_n}\|_2=\lim_n\|\et^{k_n+1}-\et^{k_n}\|_2=0.\end{equation}
Similar results hold when $\d=0,\eta\neq 0$ and $\d=1,\eta\neq 0$.
{\bf Remark:} Such properties do not guarantee the convergence of
the sequences. For example, the sequence $(u_n)_n$ defined by
$u_n=\sin(\log(n+1))$ verifies $\sum_{n=0}^{+\infty}(u_{n+1}-u_n)^2<+\infty$, however it does
not converge.

\subsubsection{Weak convergence of $(\e^{k_n+p})_n$ and $(\et^{k_n+p})_n$}
Let $\Check{\e}$ be a test function in $\mathbb{L}^2([0,T];\mathbb{R})$. From:
\begin{equation}\nonumber <\Check{\e},\e^{k_n+1}>=<\Check{\e},\e^{k_n+1}-\e^{k_n}>+<\Check{\e},\e^{k_n}>,\end{equation}
one can easily prove that $(\e^{k_n+1})_n$ weakly converges to
$\e$ too. By the same way, $(\e^{k_n+p})_n$ and $(\et^{k_n+p})_n$ also
weakly converges to $\e$.

\subsection{Strong convergence of $(\e_{k_n})_n$}
\subsubsection{Strong convergence of $(\p^{k_n})_n$, $(\c^{k_n-1})_n$ and
  $(\c^{k_n})_n$}
Since we have proven that $(\e_{k_n})_n$ and $(\et_{k_n})_n$ weakly
converge in $\mathbb{L}^2([0,T];\mathbb{R})$, hence in $\mathbb{L}^1([0,T];\mathbb{R})$, we can use theorem 3.6 of (\cite{ball}), which implies that $\p^{k_n}$
strongly converges in $\mathcal{C}([0,T];\mathbb{L}^2(\Omega,\mathbb{C}))$ to the state $\p$
associated to $\e$. One can also easily
adapt the proof of this theorem in order to obtain that $(\c^{k_n-1})_n$
and $(\c^{k_n})_n$ also strongly converge in $\mathcal{C}([0,T];\mathbb{L}^2(\Omega,\mathbb{C}))$
to the adjoint state associated to $\e$ and $\p(T)$.
\subsubsection{Strong convergence of $(\e^{k_n})_n$}
The strong convergences of $(\p^{k_n})_n$, $(\c^{k_n-1})_n$ and $(\c^{k_n})_n$ implies the strong
convergence of $(\frac{\d}{\a}\textrm{Im}\langle \c^{k_n-1}|\m|\p^{k_n}\rangle )_n$ and
$(\frac{\eta}{\a}\textrm{Im}\langle \c^{k_n}|\m|\p^{k_n}\rangle)_n$ in
$\mathcal{C}([0,T];\mathbb{R})$. According to the definitions (\ref{monotonic}) and
(\ref{monotonic2}), we can now write $(\e^{k_n})_n$ as follows:
\begin{equation}\nonumber
\e^{k_n+1}=\underbrace{(1-\delta)(1-\eta)}_{\lambda}\e^{k_n}+u_n,
\end{equation}
where $(u_n)_n$ strongly converges. Let $e$ denote a positive real
number. Since $|\lambda|<1$, there exists an integer $j_0$ such that
$|\lambda^{j_0}|<e$. Let us write then:
\begin{equation}\label{decompo}
\e^{k_n}=\e^0+\sum_{j=0}^{j_0-1}\lambda^ju_{k_n-j-1}+\lambda^{j_0}\sum_{j=0}^{k_n-j_0}\lambda^ju_{k_n-j-1}.
\end{equation}
The second term of (\ref{decompo}) is a finite sum of strongly
convergent contributions and its third term can be bounded above by
$e\|u\|_2\frac{1}{1-\lambda}$, which ends the proof of the
strong convergence of $(\e^{k_n})_n$. The strong limit is necessarily $\e$. Passing to the limit in (\ref{monotonic}), we conclude that $\e$ is a
critical point of $J$.\\
A similar proof can be done to prove that $(\et^{k_n})_n$ strongly
converges to $\e$.\\
It thus has been proven that every
weakly convergent subsequence of $(\e^k)_k$ strongly converge in
$\mathcal{C}([0,T];\mathbb{R})$ to a critical point of $J$.

\section{PROPERTIES OF THE LIMIT SET}
This section is devoted to the study of the limit set of the
sequence $(\e^k)_k$ which will be denoted by $A$. From now on we will suppose that this
set contains at least two points\footnote{Of course $A$ depends on the initial guess $\e^0$ for the definition of the sequence $(\e^k)_k$ (i.e. $A=A(\e^0)$ even though we shall skip this dependency in all what follows).}.\\
\subsection{First properties}
From theorem \ref{borne}, one deduces that $A\subset B(0,M)$, where
$M$ is defined in (\ref{defM}) and $B(0,M)$ stands for the ball of radius $M$ of $\mathbb{L}^2([0,T];\mathbb{R})$.
According to the definition of $A$, the results of the latest section prove that $A$ is
a subset of the set of critical points of $J$. Finally, thanks to the monotonic
property (\ref{monoprop}) of the sequence $(\e^k)_k$, we deduce that $J$
is constant on $A$.

\subsection{Compactness}
Let us now prove a first topological property.
\begin{lemma}
The set $A$ is compact.
\end{lemma}
\begin{proof}
Let $(\e_{\infty}^n)_n$ denote a sequence of $A$. We can associate to
this sequence a subsequence $(\e^{k_n})_n$ of $(\e^k)_k$ such that
$\|\e_{\infty}^n-\e^{k_n}\|_2<\frac{1}{n}$. According to the previous
results we can extract from $(\e^{k_n})_n$ a strongly convergent
subsequence $(\e^{k_{n'}})_{n'}$. Let $\e^*$ denote the limit of this
latter. Thus, the sequence $(\e_{\infty}^{n'})_{n'}$ strongly
converges to $\e^*$ that is a point of $A$.
\end{proof}

\subsection{e-Strings in $A$}
Consider a general metric space $(E,d)$, $(x,y)\in E^2$ and $e$ a positive
real number. We
call $e$-string between $x$ and $y$ a finite sequence $z_1,...,z_N$ of
point of $E$ such that:
\begin{itemize}
\item $z_1=x$,
\item $z_N=y$,
\item $\forall k\in [1,N-1],\ d(z^{k+1},z^k)<e$.
\end{itemize}
Then the set $A$ has the following topological property.\\
\begin{lemma}
 For any $(\e_{\infty},\e'_{\infty})\in A^2$ and any $e>0$, there
exits an $e$-string in $A$ between $\e_{\infty}$ and $\e'_{\infty}$.
\end{lemma}
\begin{proof}
As a compact set, there exist $N_0$
open balls of radius $\frac{e}{4}$
covering $A$.
By the definition of $A$ and  (\ref{limite}), there exists an infinity
of $K>0$ for which $l_K=\e_{\infty},\e^{K},...,\e^{K+N(K)},\e'_{\infty}$ is an $e$-string. From $l_K$, one can then build another $e$-string
$l'_K=\e^{K,1},\e^{K,2},...,\e^{K,N_0}$. Indeed, if $N_0>N(K)$, define
$l'_K$ by:
$$l'_K=l_K,\underbrace{\e^{K+N(K)},...,\e^{K+N(K)}}_{N_0-N(K) \ terms},$$
and if $N_0>N(K)$, one can remove $N(K)-N_0$ terms of $l_K$ while
keeping the $e$-string properties. \\ For each $i,\ 1\leq i \leq N_0$,
let us extract from $(\e^{K,i})_K$ a strongly convergent subsequence of
$(\e^k)_k$. The limits obtained are an $e$-string in $A$.
\end{proof}

\subsection{Connexity}
The previous result leads to the following theorem.
\begin{theorem}
The set $A$ is connex.
\end{theorem}
\begin{proof}
Suppose there exist two closed subsets of $A$, denoted by $A_1$ and $A_2$, such
that $A=A_1\cup A_2$ and $A_1\cap A_2=\emptyset$. Because of the existence of
$e$-strings for every $e$, we deduce that the distance between $A_1$
and $A_2$ is equal to 0. Since $A$ is compact, this is in contradiction with $A_1\cap A_2=\emptyset$.
\end{proof}

\subsection{Summary}
It has been proven that the limit points of a sequence obtained by a
monotonic scheme are a compact and connex set of critical points of
$J$. Note that if this set is reduced to one point, the compactness of
the sequence implies its
convergence.

\section{VARIATIONAL ANALYSIS AND PARTICULAR CASE}
Let us focus now on the scheme obtained for $\delta=1$ and $\eta=0$,
which corresponds to the Krotov formulation (as in \cite{tannor}).
We will estimate the variations of $\p$ and $\chi$ with respect
to $\e$. The results obtained will enable us to prove the convergence for
large values of the parameter $\alpha$.\\
The above defined set $A$ is still considered to contain at least two points.
\subsection{Estimates }
Let $\e$ and $\e'$ be two points of $A$, $\p$ and $\p'$ the
corresponding states given by $(\ref{el1})$ and $\c$ and $\c'$ the
corresponding adjoint states solution of $(\ref{el3})$. Consider $(\ref{el1})$
written in integrated form, for $\p$ and $\p'$:
\begin{eqnarray}
\p(t)&=&e^{-iHt}\psi_0+\int_0^te^{-iH(t-s)}\e(s) i \m \p(s)ds,\nonumber\\
\p'(t)&=&e^{-iHt}\psi_0+\int_0^te^{-iH(t-s)}\e'(s) i \m \p'(s)ds.\nonumber
\end{eqnarray}
Let us introduce the notations $\delta \p(t)=\p(t)-\p'(t)$, $\delta \c(t)=\c(t)-\c'(t)$ and $\delta \e(t)=\e(t)-\e'(t)$, we then have:  
\begin{eqnarray}\nonumber\label{pregron}
\delta \p (t)&=&\int_0^t e^{-iH(t-s)}\delta \e(s) i\mu\p (s)ds\\&&+\int_0^t
e^{-iH(t-s)}\e'(s) i\mu\delta \p (s)ds.
\end{eqnarray}
Since the operator $e^{-iHt}$ is unitary, we deduce that:
\begin{eqnarray}\nonumber
\!\!\!\!||\!\!\int_0^t\!\!\!\! e^{-iH(t-s)}\delta \e(s) i\mu\p (s)ds||\!\!&\!\!\!\!<\!\!\!\!&\!\!||\mu||_*T\|\delta
  \e\|_2,\\
\!\!\!\!||\!\!\int_0^t\!\!\!\!
e^{-iH(t-s)}\e'(s) i\mu\delta \p (s)ds||\!\!&\!\!\!\!<\!\!\!\!&\!\!M||\mu||_*\int_0^t\!\!\!\!||\delta \p
  (s)||ds,\nonumber
\end{eqnarray}
where $M$ has been defined in (\ref{defM}). From Gronwall's
lemma applied to (\ref{pregron}), we obtain:
\begin{equation}\label{estpsi}
||\delta \p(t)||\leq||\mu||_*Te^{T||\mu||_*M}\|\delta \e\|_1,
\end{equation}
where $\|.\|_1$ represents the norm of $\mathbb{L}^1([0,T];\mathbb{R})$.
A similar computation for the adjoint state leads to:
\begin{equation}\label{estchi}
||\delta \c(t)||\leq ||O||_*||\mu||_*T(1+e^{T||\mu||_*M})e^{T||\mu||_* M}\|\delta \e\|_1.
\end{equation}

\subsection{Convergence}
Since $\e$ and $\e'$ are critical points of
$J$, the two following equalities hold:
\begin{eqnarray}\nonumber
\alpha \varepsilon(t) &=& - \textrm{Im}  \langle \chi(t) | \mu | \psi(t)
\rangle ,\\
\alpha \varepsilon'(t) &=& - \textrm{Im}  \langle \chi'(t) | \mu | \psi'(t) \rangle .\nonumber
\end{eqnarray}
The difference of these two equalities yields:
\begin{equation}\nonumber
\alpha \delta \varepsilon(t) =- \textrm{Im}  (\langle \delta \chi | \mu | \psi
\rangle  (t)+\langle \chi | \mu | \delta \psi
\rangle  (t)).
\end{equation}
From (\ref{estpsi},\ref{estchi}) we have:
\begin{equation}\label{enfin}
\alpha \|\delta \e\|_1 \leq ||O||_*||\mu||_*^2T^2
(1+e^{T||\mu||_*M})e^{2T||\mu||_* M} \|\delta \e\|_1.
\end{equation}
Thus we get the following result:\\
\begin{theorem} The monotonic scheme defined by (\ref{monotonic})-(\ref{monotonic4}), $\delta=1$,
$\eta=0$  strongly converges in
$\mathbb{L}^2([0,T];\mathbb{R})$ under the assumption that: \begin{equation}\nonumber \alpha > ||O||_*||\mu||_*^2T^2
(1+e^{T||\mu||_*M})e^{2T||\mu||_* M}.\end{equation}\end{theorem}\vspace{.5cm}
\begin{proof}Suppose that the monotonic scheme does not converge, then there exists at least two distinct
points $\e$ and $\e'$. Using the above notations, the
equation (\ref{enfin}) holds in this case. Since $\delta \e \neq 0$,
we reach a contradiction. \end{proof}

\section{CONCLUSION}

It has been proven that the sequences provided by monotonic schemes
are compact and that the set of their limit points is compact and connex. It has been shown that
this set reduces to one point (i.e. the algorithm strongly converges)
for a large laser fluence penalty parameter $\alpha$. We refer the reader to
\cite{these} for a more detailed presentation of this topic.

\section{ACKNOWLEDGMENTS}
It is a pleasure to acknowledge helpful discussions that we had on
this topic with
Y. Maday (Jacques-Louis Lions laboratory, Paris) and G. Turinici
(INRIA, Rocquencourt and CERMICS-ENPC, Marne-la-Vall{\'e}e).






\begin{thebibliography}{99}




\bibitem{br}
H. Rabitz, Shaped laser pulses as reagents, {\it Science}, 299, 2003,
pp 525-526.

\bibitem{book}
Herschel Rabitz, Gabriel Turinici and Eric Brown, {\it Control of quantum dynamics: Concepts, procedures and future
  prospects}, \newblock In Ph.~G. Ciarlet, editor, {\it Computational Chemistry, Special
  Volume (C. Le Bris Editor) of Handbook of Numerical Analysis, vol
  X}, Elsevier Science B.V.; 2003, pp 833-887.

\bibitem{judson}
R.~S. Judson and H.~Rabitz, Teaching lasers to control molecules, {\it
  Phys. Rev. Lett.}, 68, 1992, pp 1500-1503.

\bibitem{shi1}
S. Shi, A. Woody, and H. Rabitz, Optimal Control of Selective
Vibrational Excitation in Harmonic Linear Chain Molecules,  {\it
  J. Chem. Phys.}, 88, 1988, pp 6870-6883.

\bibitem{scienceref1}
R.~J. Levis, G.~Menkir and H.~Rabitz, Selective bond dissociation and
  rearrangement with optimally tailored, strong field laser pulses,
  {\it Science}, 292, 2001, pp 709-713.

\bibitem{hbref1}
  A.~Assion,  T.~Baumert,
  M.~Bergt,
  T.~Brixner,
  B.~Kiefer,
  V.~Seyfried,
  M.~Strehle and
  G.~Gerber, Control of chemical reactions by feedback-optimized phase-shaped
  femtosecond laser pulses, {\it Science}, 282, 1998, pp 919-922.

\bibitem{hbref2}
  M.~Bergt,
  T.~Brixner,
  B.~Kiefer,
  M.~Strehle and
  G.~Gerber, Controlling the femto-chemistry of $Fe(CO)_5$, {\it
  J. Phys. Chem. A.}, 103, 1999, pp 10381-10387.

\bibitem{hbref4}
  T.~Weinacht,
  J.~Ahn and
  P.~Bucksbaum, Controlling the shape of a quantum wavefunction, {\it
  Nature}, 397, 1999, pp 233-235.

\bibitem{hbref5}
  C.~Bardeen,
  V.~V. Yakovlev,
  K.~R. Wilson,
  S.~D. Carpenter,
  P.~M. Weber and
  W.~S. Warren, Feedback quantum control of molecular electronic
  population transfer, {\it Chem. Phys. Lett.}, 280, 1997, pp 151-158.

\bibitem{hbref6}
  C.~J. Bardeen,
  V.~V. Yakovlev,
  J.~A. Squier
  and K.~R.  Wilson, Quantum control of population transfer in green fluorescent protein by using
  chirped femtosecond pulses, {\it J. Am. Chem. Soc.}, 120, 1998, pp 13023-13027.

\bibitem{zhu}
  W.~Zhu and H.~Rabitz, A rapid monotonically convergent iteration algorithm for quantum optimal
  control over the expectation value of a positive definite operator,
  {\it J. Chem. Phys.}, 109, 1998, pp 385-391.



\bibitem{tannor}
  D.~Tannor,
  V.~Kazakov and
  V.~Orlov, {\it Control of photochemical branching: Novel procedures for finding optimal
  pulses and global upper bounds} In
  {\it Time Dependent Quantum Molecular Dynamics}, edited
  by Broeckhove J. and
  Lathouwers L. Plenum; 1992, pp 347-360.


\bibitem{jcp}
  Y.~Maday and G.~Turinici, New formulations of monotonically
  convergent quantum control algorithms,
  {\it J. Chem. Phys.}, 118 (18), 2003, pp 8191-8196.
\newpage
\bibitem{seville}
Y.~Maday, J.~Salomon and G.~Turinici, "Discretely
monotonically convergent algorithm in quantum control", {\it
  Proceedings of the LHMNLC03 IFAC conference}, Sevilla, Spain, April 2003, pp 321-323.

\bibitem{grivopoulos}S.~Grivopoulos and B.~Bamieh, "Iterative
  algorithms for optimal control of quantum systems", {\it Proceedings
    of the 41st IEEE Conference on Decision and Control}, Las Vegas, Nevada, December 2002, pp 2687-2691.

\bibitem{hornung}
  T.~Hornung,
  M.~Motzkus
  and R.~de~Vivie-Riedle, Adapting optimal control theory and using learning loops to provide
  experimentally feasible shaping mask patterns, {\it J. Chem.Phys.},
  115, 2001, pp 3105-3111.

\bibitem{innsbruck}
J.~Salomon and G.~Turinici, "Control of molecular orientation
and alignment by monotonic schemes", {\it Proceedings of the 24-th
  IASTED International Conference on modelling, identification and
  control}, Innsbruck, Austria, February 2005, 457-187.

\bibitem{ball}
J.M. Ball, J.E. Marsden an M. Slemrod, Controllability for distributed
bilinear systems, {\it SIAM J. Control and Optimization}, 20 (4),
1982, pp 575-597.










\bibitem{these}
 J.~Salomon, Ph.D. Thesis, \it in progress.\rm












\end{thebibliography}
\end{document}